\documentclass{elsart}
\usepackage{amssymb}
\usepackage{amsfonts}

\usepackage{graphicx}

\usepackage{epsfig}

\begin{document}
\begin{frontmatter}

\title{Offdiagonal Complexity: A 
computationally quick
complexity measure
 for graphs and networks
\\[1ex]
{\normalsize\sl (Physica A, in print)}
}
\author{Jens Christian Claussen}
\address{Institut f\"ur Theoretische Physik und Astrophysik,\\
Universit\"at Kiel, Leibnizstra{\ss}e 15, 24098 Kiel, Germany
 \\
phone ++49-431-880-4096
\\ \tt claussen@theo-physik.uni-kiel.de
}
\date{October 16, 2004; revised August 11, 2006}
\begin{abstract}
A vast variety of biological, social, and economical networks shows
topologies drastically differing from random graphs; yet the quantitative
characterization remains unsatisfactory from a conceptual point of view.
Motivated from the discussion of small scale-free networks, a biased link
distribution entropy is defined, which takes an extremum for a power law
distribution. This approach is extended to the node-node link
cross-distribution, whose nondiagonal elements characterize the graph
structure beyond link distribution, cluster coefficient and average path
length. From here a simple (and computationally cheap) complexity measure
can be defined.
This Offdiagonal Complexity (OdC) is proposed as a novel measure to characterize the 
complexity of an undirected graph, or network.
While both for regular lattices and fully connected 
networks OdC is zero, it takes a moderately low value for a random graph
and shows high values for apparently complex 
structures as scale-free networks and hierarchical trees. 
The Offdiagonal Complexity apporach is applied to the Helicobacter pylori
protein interaction network and randomly rewired surrogates.
\end{abstract}
\end{frontmatter}

\vspace{0.4in}

\noindent
\section{Introduction}
While random graph theory and scale-free network research 
know a set of standard measures to quantify their 
properties,
the question of {\sl complexity} of a graph
still is  in its infancies. 
A `blind' application of other complexity measures 
(as for binary sequences or computer programs)
does not account for the special properties
shared by graphs and especially scale-free graphs.
Moreover, some known complexity measures themselves
have a high computational complexity.
\\
Since a series of seminal papers  
(Watts \& Strogatz \cite{wattsstrogatz}, 
Barabasi \& Albert
 \cite{barabasialbert}
\cite{barabasialbert,albertbarabasi},
Newman \cite{newman}, 
Dorogovtsev \& Mendes  \cite{doromend}) 
since 1999 (see also \cite{bornschu} for an overview),
small-world and
scale-free networks are a hot topic of investigation
in a broad range of systems and disciplines.
Metabolic and other 
biological networks,
collaboration networks, www, internet, etc.,
have in common
that the distribution of link degrees follows a 
power law, thus has no inherent scale.
Such networks are termed `scale-free networks'.
Compared to random graphs,
which have 
a Poisson link distribution and thus
a characteristic
scale, they share a lot of different properties,
especially 
a high clustering coefficient, and
a short average path length. 
\\
Mathematically, a graph (or synonymously in this context,
a network) is defined by a (nonempty) set of nodes, a set of
edges (or links), and a map that assigns two nodes
(the ``end nodes'' of a link) to each link.
In a computer, a graph may be represented either
by a list of links, represented by the pairs of nodes,
or equivalently, by its adjacency matrix $a_{ij}$ whose
entries are 1 (0) if nodes $i,j$ are connected (disconnected).
Useful generalizations are weighted graphs, where the 
restriction of $a_{ij}$ is relaxed from binary values
to (unsually nonnegative) integer or real values 
(e.g.\ resistor values, travel distances, interaction coupling),
and directed graphs, where $a_{ij}$ no longer needs
to be symmetric, and the link from $i$ to $j$ and 
the link from $j$ to $i$ can exist independently
(e.g.\ links between webpages, or scientific citations).
\\
Here the discussion will be kept limited to binary 
undirected graphs, like an acquaintancy network
or a railway network as shown below.
In the following sections the link (degree) distribution 
and the next order cross-distribution
are investigated
and taken as a basis for a complexity measure.

\vspace{0.4in}

\section{Other complexity measures}
\vspace*{-.8ex}
For text strings (as computer programs, or DNA)
there are common 
 complexity measures 
in theoretical computer science,
as {\sl Kolmogorov complexity}
(and the related {\sl  Lempel-Ziv complexity}
and  {\sl algorithmic information content} AIC)
\cite{gellmannLloyd}.
E.g., AIC is defined by the length of the shortest
program generating the string.
For random structures, thus also for random graphs, 
they indicate high complexity.
A distinction of complex structured (but still partly random)
structures from completely random ones 
usually is prohibitive for this class of measures.
For this reason, measures of {\sl effective complexity}
\cite{gellmann} have been discussed; 
usually these are defined as an entropy (or description length)
of ``a concise description of a set of the entity's regularities''
\cite{gellmann}.
Here we are mainly interested in this second class,
and straightforwardly one would try to apply 
existing measures, e.g., to the link list or to the
adjacency matrix. 
However, mathematically it is not straightforward
 to apply these text string based measures to graphs,
as there is no unique way to map 
a graph onto a text string.
For the case of hierarchical structures, which can be
represented by trees, 
Ceccatto and Huberman
quantified complexity 
from the diversity of the subtrees
\cite{ceccatohuberman}.
As natural networks typically exhibit 
an occurrence triangles and higher oder loops
in a nonneglectable way, 
other approaches have to be chosen for networks in general.

Thus one desires to use complexity measures
that are defined directly 
for graphs.
Two classical
measures are known from graph theory,
 {\sl graph thickness} 
and
 {\em coloring number}
have a low ``resolution''
(typically integer values up to 4),
and their relevance
for real networks is not clear.
Two new complexity measures 
recently have been proposed for graphs,
{\sl Medium Articulation}
\cite{wilhelm}
for weighted graphs
(as they appear in
foodwebs)
and 
a measure 
for directed graphs
by Meyer-Ortmanns
\cite{meyer}
based on the 
{\sl network motif} concept 
\cite{alon_motif}).
Unfortunately, the latter two complexity measures are 
computationally quite costly.
A computational complexity approach has been 
defined by Machta and Machta \cite{machtamachta}
as {\em computational depth} 
of an {\em ensemble of graphs}
(e.g.\ small-world, scalefree, lattice). 
It is defined as the number of
processing time steps a large parallel
computer (with unlimited number of processors)
would need to generate a {\em representative}
member of that graph ensemble.
Unlike other approaches, it does not assign single 
complexity values to each graph,
and again is nontrivial to compute.

\begin{figure}
\noindent
\noindent
\begin{flushleft}
\begin{tabular}{|c|c|l|}
\hline
&&\\[-4mm]
\epsfig{file=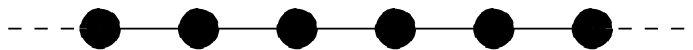,width=3.0cm}
&
\epsfig{file=gra1dp.eps,width=3.5cm}
&  
{\small 
\footnotesize
\begin{tabular}[b]{c|ccc}  
&2&$k_2$\\\hline    
2&$\bullet$&\\    
$k_1$&&          
\end{tabular}   
} 
\\
\hline
&&\\[-4mm]
\epsfig{file=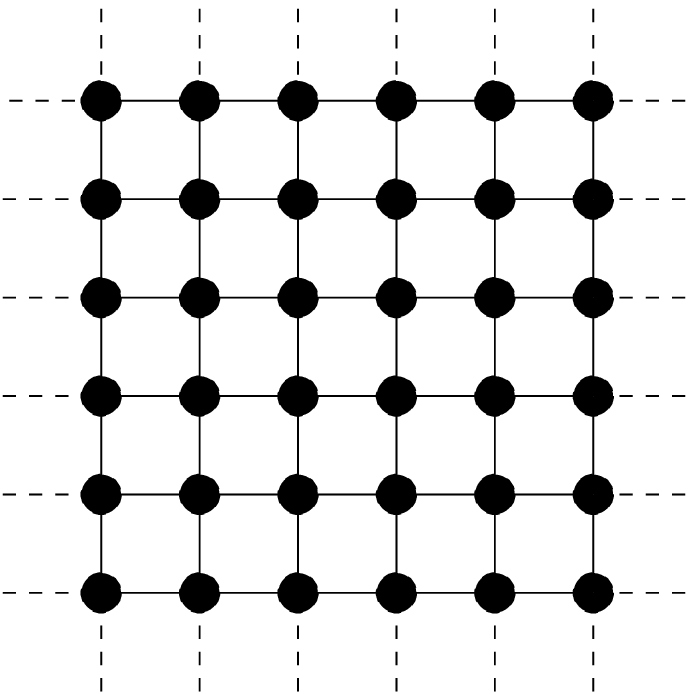,width=3.0cm}     
&
\epsfig{file=gra2dp.eps,width=3.5cm}
&
{\small 
\footnotesize
\begin{tabular}[b]{c|ccc}
&4&$k_2$\\\hline
4&$\bullet$&\\
$k_1$&&   
\end{tabular}
} 
\\
\hline
&&\\[-4mm]
\epsfig{file=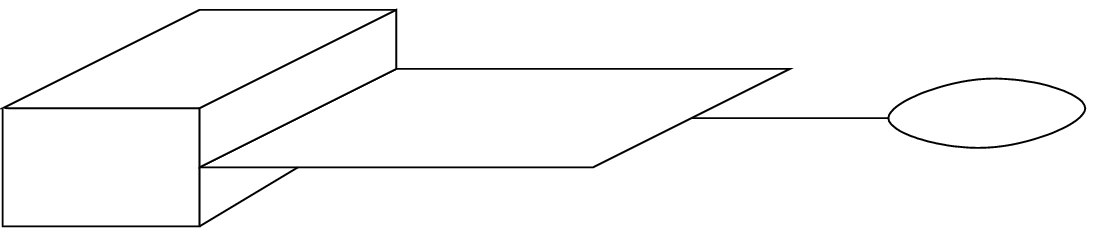,width=5.5cm}
&  
\epsfig{file=gra_boxplanestickloopp.eps,width=3.5cm}
&  
{\small 
\footnotesize
\begin{tabular}[b]{c|ccccc}
&2&4&8&$k_2$\\\hline
2&$\bullet$&0&0&\\
4&&$\bullet$&0&\\
8&&&$\bullet$&\\
$k_1$&&&&   
\end{tabular}
} 
\\
\hline
&&\\[-4mm]
\epsfig{file=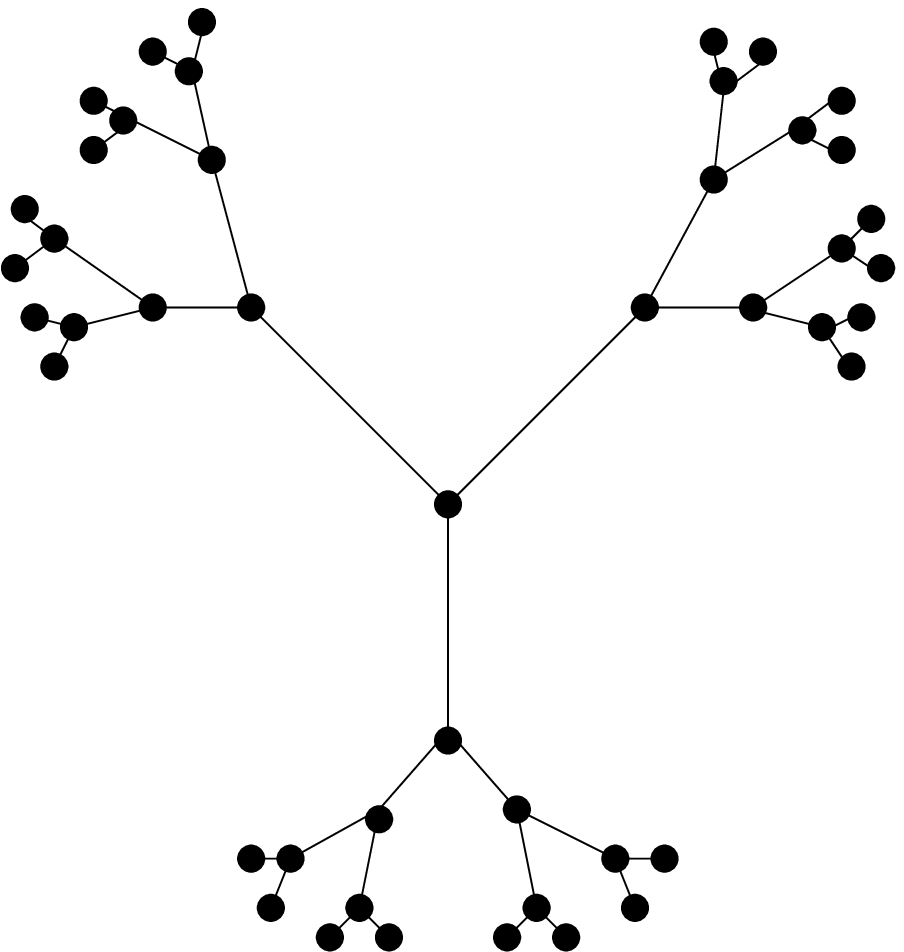,width=4.0cm}     
&  
\epsfig{file=gra_bethep.eps,width=3.5cm}
&  
{\small
\footnotesize
\begin{tabular}[b]{c|ccccc}
&1&2&3&$k_2$\\\hline
1&0&0&$\bullet$&\\  
2&&0&0&\\   
3&&&$\bullet$&\\    
$k_1$&&&&   
\end{tabular}
} 
\\
\hline
&&\\[-4mm]
\epsfig{file=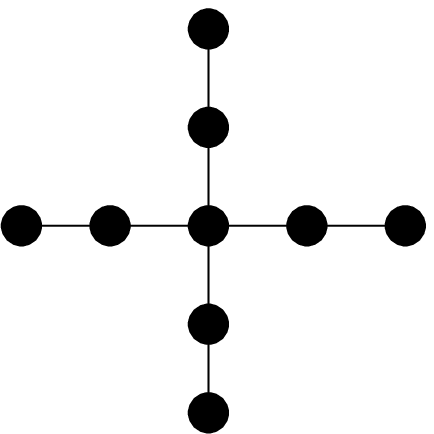,width=2.0cm}     
&  
\epsfig{file=gra_star4p.eps,width=3.5cm}
&  
{\small 
\footnotesize
\begin{tabular}[b]{c|ccccc}
&1&2&4&$k_2$\\\hline
1&0&4&0&\\
2&&0&4&\\
4&&&0&\\   
$k_1$&&&&   
\end{tabular}
} 
\\
\hline
\end{tabular}
\end{flushleft}
\caption{Small non-complex networks:
These networks are large, not complex, and not scale-free. A single 
entry or a single diagonal with nonzero entries indicates low complexity.
Shown are a regular lattice in 1D and 2D (top) and 
a Bethe lattice and a star graph (bottom)
The third example (middle) is the 
box-plane-stick-loop concatenation of
different-dimensional finite lattices,
widely used as data analysis test set.
\label{fignoncomplex}}
\end{figure}

\begin{figure}
\begin{flushleft}
\footnotesize
\begin{tabular}[b]{|c|c|l|}
\hline
&&\\[-2mm]
\epsfig{file=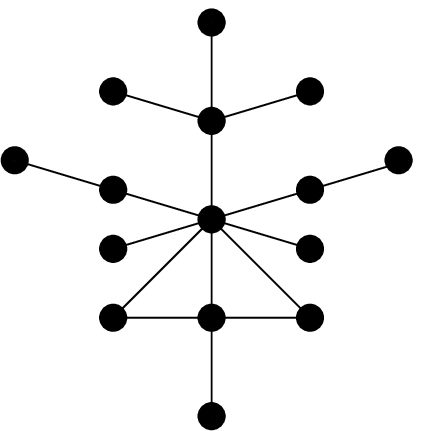,width=3.2cm}
&
\epsfig{file=gra_com0p.eps,width=3.8cm}
&  
{\small 
\footnotesize
\begin{tabular}[b]{c|ccccc}
&1&2&4&8&$k_2$\\\hline
1&0&2&4&2&\\
2&&0&2&4&\\
4&&&0&2&\\
8&&&&0&\\
$k_1$&&&&&
\end{tabular}
}
\\
\hline
&&\\[-2mm]
\epsfig{file=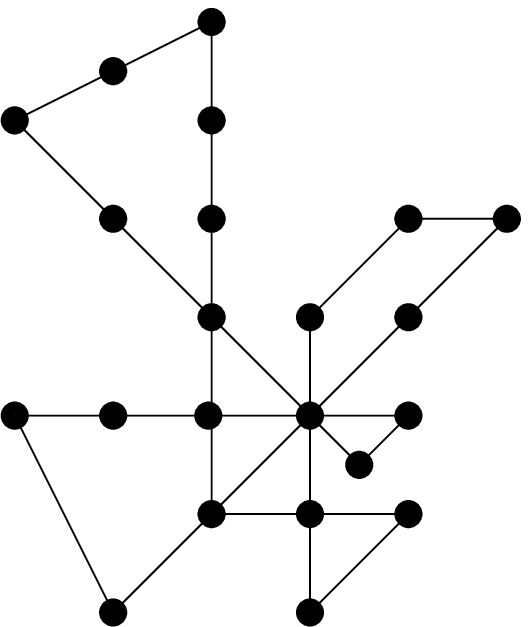,width=3.2cm}   
&  
\epsfig{file=gra_com1p.eps,width=3.8cm}
&  
{\small 
\footnotesize
\begin{tabular}[b]{c|cccc}
& 2&4&8&$k_2$\\\hline
2&11&6&4&\\
4&&3&4&\\
8&&&0&\\
$k_1$&&&&
\end{tabular}
}
\\[2mm]
\hline
&&\\[-2mm]
\epsfig{file=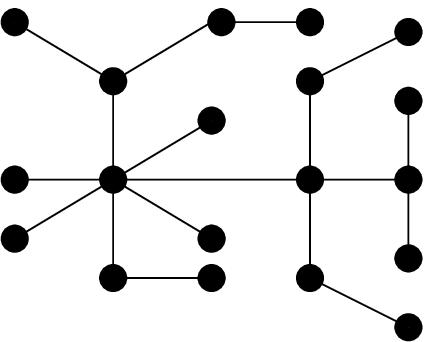,width=3.2cm}
&  
\epsfig{file=gra_com2p.eps,width=3.8cm}
&  
{\small 
\footnotesize
\scriptsize
\begin{tabular}[b]{c|cccccccc}
&1&2&3&4&5&6&7&$k_2$\\\hline  
1&0&4&3&0&0&0&4&\\
2&&0&1&2&0&0&1&\\
3&&&0&1&0&0&1&\\
4&&&&0&0&0&1&\\
&&&&&0&0&0&\\
&&&&&&0&0&\\
$k_1$&&&&&&&0&\\

\end{tabular}
}
\\
\hline
&&\\[-2mm]
\epsfig{file=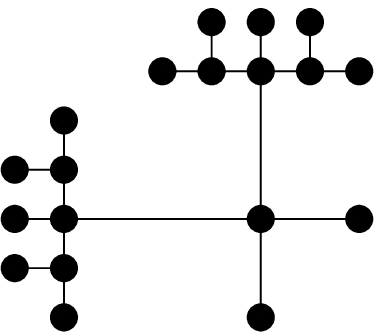,width=3.2cm}     
&  
\epsfig{file=gra_com3p.eps,width=3.8cm}
&  
{\small 
\footnotesize
\begin{tabular}[b]{c|ccccc}
&1&2&3&4&$k_2$\\\hline
1&0&8&0&4&\\
2&&0&0&4&\\
3&&&0&0&\\
4&&&&2&\\
$k_1$&&&&&
\end{tabular}
}
\\
\hline
\end{tabular}
\normalsize
\end{flushleft}
\caption{Small complex networks:
A striking observation is that
entries are quite evenly spread on the offdiagonals.
Can this be used to \mbox{define} a complexity measure?
\label{figcomplex}}
\end{figure}

Following \cite{gellmann}, 
an  especially desired property of a complexity measure 
should be the ability 
to distinguish nonrandom complex structures from 
both pure randomness and regular structures as lattices.
In this instance, the effective complexity and the
Machta approach fulfill this prerequisites perfectly,
but up to today no simple method is available to
compute them.
Hence,  a 
{\sl simpler 
estimator} 
of graph complexity is
desired,
and one possible approach, the
Offdiagonal Complexity,
is proposed here.
It is motivated by a striking observation 
on the node-node link correlation matrices
of complex networks
\cite{claussenddhs03}, 
namely that
entries are more evenly spread among the offdiagonals,
compared to both regular lattices and random graphs
(see Figs.\ \ref{fignoncomplex}  and \ref{figcomplex} for a comparison).
This can be used to \mbox{define} 
a complexity measure, 
 for undirected
graphs
 \cite{claussenddhs03,claussenECMTB}.

\vspace{0.4in}

This article is organized as follows.
 In Sec.\ \ref{sec_motivationODC} the approach is motivated from link entropies and node-node correlations.
In Sec.\ \ref{sec_ODCdef}
OdC is defined.
Section
\ref{sec_helico}
investigates the application of OdC to 
a protein interaction network, compared with
randomized surrogates.

\vspace{0.4in}

\section{Motivation of OdC\label{sec_motivationODC}}
\vspace*{-2ex}
\subsection{Node degree correlations: Methods of classical statistics}
\vspace*{-.8ex}
A straightforward mathematical approach to study 
node-node link correlations,
i.e.\ correlations between 
degrees of pairs of nodes,
is to use
 rank correlation methods 
\cite{kendall90}
from classical statistics
to analyze the link distributions.

Two common 
rank correlation
  methods 
can be described as follows.
One considers a list of {\sl rank numbers} of link numbers
(node degrees).
For each of the two graphs (A and B)
to be compared, there is a (ordered) 
list of 
link numbers 
$(k_1,k_2,\cdots k_N)=$
\mbox{5 2 2 1 1 1},
and one assigns a rank number to each link,
$(r^{\rm A}_1, r^{\rm A}_1, \ldots r^{\rm A}_N)=$
(1 2.5 2.5 5 5 5).
Hereby the identical second and third ranks are replaced by the
(noninteger) average value; as node degrees are highly degenerate,
this will occur frequently.

\noindent
Then the Kendall tau coefficient is defined as
$ 
t = \displaystyle\frac{2\sum_{ij} \sigma_{ij}}{n(n-1)},
$
where $\sigma_{ij}=\pm 1$, 
if pairs of elements $(i,j)$ 
are ranked in both lists equally (resp.\ non equally),
\\$\sigma_{ij}={\rm sgn}(r^{\rm A}_i-r^{\rm A}_j)  \cdot{\rm sgn}(r^{\rm B}_i-r^{\rm B}_j)$.
Its apparent drawbacks within this context are
the required  costly computations ($n^2$),
and it seems to be 
analytically 
not easy to handle, as one must have the nodes sorted
by their degree, for each member of (e.g.) an ensemble average.

The second main rank correlation method is Spearman's rho,
defined by
$
r_s 
=
 1 - 
\displaystyle\frac{6\sum_i d_i^2}{n^3-n}, 
$ where $
d_i 
= 
r^{\rm A}_i-r^{\rm B}_i.
$
--- Some of its properties are:
\\$r_s=+1$ ~~ for {\sl identical} rank lists
{ \begin{tabular}{ccc}1&2&3\\1&2&3\end{tabular} }
\\$r_s=-1$ ~~ for {\sl counter-sequenced} rank lists 
{ \begin{tabular}{ccc}1&2&3\\3&2&1\end{tabular} }\\
\\$r_s=+\frac{1}{2}$ ~~ if a sequence is constant $=\frac{n(n-1)}{2}$.
(One might wonder why not  $r_s=0$ holds here.
However for $n=3$ always $r_s \neq 0$ holds; but the 
average over all possible rank lists vanishes, $\langle r_s\rangle=0$.)

In general, 
rank correlation methods 
are {\sl not appropriate} for a {\sl high degeneracy},
i.e. a large number of nodes with the same number of links.

Thus, it is desired to formulate 
other measures that can estimate the complexity
of a graph from 
correlation information of pairs of nodes.
The approach of this paper
is to define an entropy-type measure.
To motivate the ansatz, 
the problem of binning 
and the definition of a link entropy 
is discussed in the next section.

\clearpage
\subsection{Fit of sparse power-law  distributions} \label{app_fit}
\begin{figure}[htbp]
\noindent
\hspace*{.5cm}
\epsfig{file=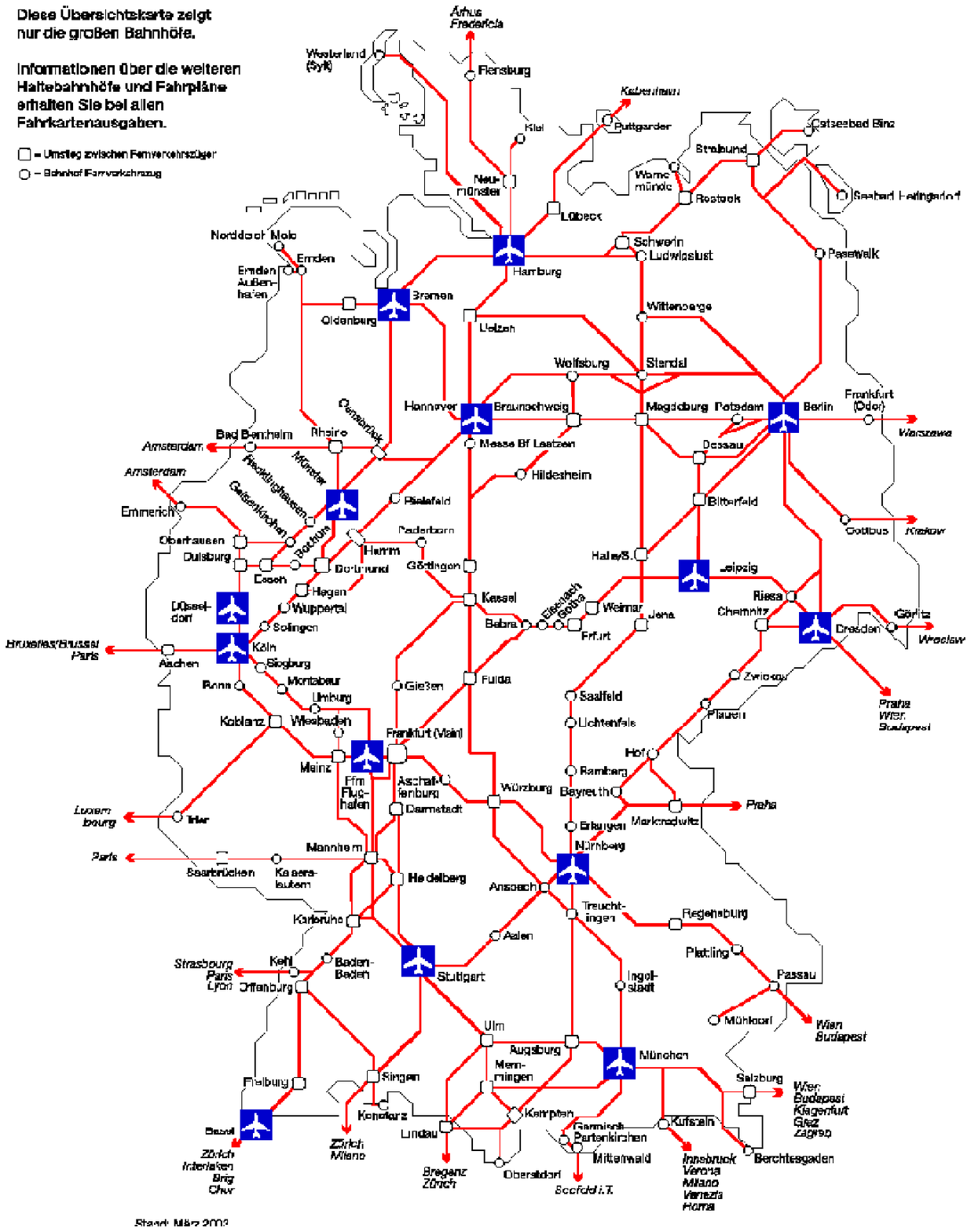,height=8.0cm}
\hspace*{1.0cm}
\raisebox{4.0cm}{{\small
\footnotesize
\begin{tabular}{|r|c|}\hline Node degree&$\#$ of nodes\\
\hline 3&20\\4&11\\5&6\\6&2\\7&1\\8&0\\9&0\\10&1\\11&0\\12&0\\13&1\\\hline
\end{tabular}
}}
\caption{Example of a small network:
The Intercity railway (plus flyway) network in Germany
approximatively shows a scale-free link distribution
(see Figs.\ \ref{fig_bin} and \ref{fig_entropy}).
\label{fig_bahn}}
\noindent
\mbox{}
\\[14mm]  
\epsfig{file=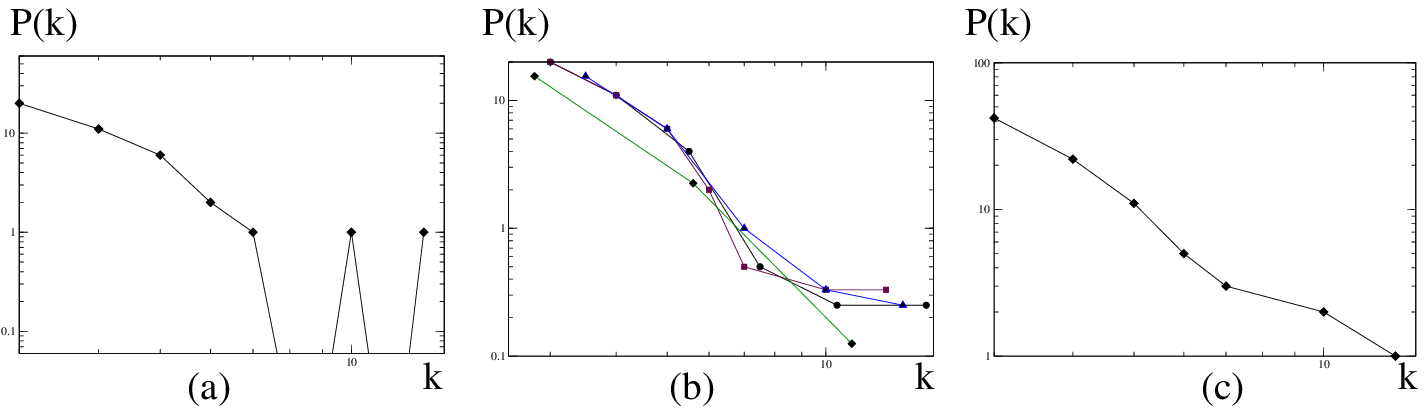,angle=0,width=0.99\textwidth}
\caption{
(a) Problem of zeroes (see text).
(b) Result of different binnings 
depending on parameters
$c$ and $n_{\rm max}$.
(c) The Integrated density is defined free of parameters.
\label{fig_bin}}
\noindent
\centerline{\epsfig{file=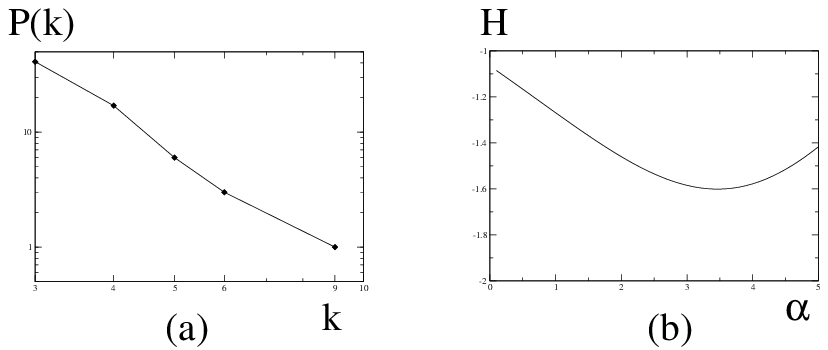,width=0.76\columnwidth}}
\caption{(a) The biased entropy of the distribution
shows an extremum with respect to the exponent $\alpha$ 
(b).
From here, we have 
a parameter-free estimation of  $\alpha$.
\label{fig_entropy}}
\end{figure}

The fit of sparse distributions by binning
has to cope with the problem of zeroes and with the 
effect of arbitrariness of the choice of interval
length and position.
As an example we consider the link distribution
of
a traffic
network 
\cite{bahn_fig}
(see Fig.~\ref{fig_bahn}).

As intervals have to be chosen so that no zeroes occur
($-\infty$ in log scale), 
one has the choice between different `tricks' (influencing the fit):
(i) irregular intervals: choice influences fit,
or 
(ii) regular intervals
$
n_{\rm max} \cdot \sqrt{2} \ln( c  \cdot \exp(k)),
$
however they
imply a severe reduction of the number of intervals.
Even the two remaining parameters influence the result
(esp.\ for large link numbers):
(see Fig.~\ref{fig_bin}b).
A moderately `clean' method is to place the entry with 
largest link number in the middle of that interval.
A parameter-free approach is the
integrated density.
For a power law density with exponent
 $\alpha>1$, one has
\begin{eqnarray}\nonumber
\int_x^\infty {\rm d}k \;\;  k^{-\alpha}= \frac{x^{1-\alpha}}{\alpha - 1},
\end{eqnarray}
Instead of the density itself, the integrated density can be fitted
(see Fig.~\ref{fig_bin}c).
For exact results, a discretization correction $c_n^\alpha$ is necessary:
$
c_n^\alpha = \frac{
\sum_{k=n}^\infty k^\alpha}{
\int_n^\infty 
 {\rm d}k\;\;  k^{-\alpha}}.
$
Alternatively, from
$\sum_{k=n}^\infty k \cdot p(k)$ one gets a plot with the same slope as
$p(k)$ itself.

\clearpage
\noindent 
\subsection{Entropy of the link distribution}
\vspace*{-.5ex}
As demonstrated in sec.~\ref{app_fit},
the estimation of the scale exponent from a 
measured distribution by binning has inherent
degrees of freedom; this can be overcome by a fit of the
integrated density.
To estimate the entropy of 
a distribution ($\neq$ density) with sampling gaps
however leads to underestimation
{(Grassberger \cite{grassberger_samp})}.
A straightforwardly defined link distribution entropy
 $H= - \sum_k {p_k} \ln p_k$ 
becomes extremal for the  {\sl  equidistribution} 
(and not for a power law).
Power law candidate distribution are usually 
logarithmically binned.
However, for a power law one obtains a
distribution with linear decay (in the binned 
log-log space, 
as in Fig.~\ref{fig_bin}b,c),
and not an equidistribution, 
and again $H$ not
maximal.

This problem is solved by defining a 
``Biased  Link Entropy''
 (showing an extremum 
w.r.t.\
$\alpha$, see Fig.\ \ref{fig_entropy};
the transformed density is the equidistribution for proper choice 
of $\alpha$).
With the necessary 
normalization $N(\alpha)=\sum_kk^\alpha p_k/\delta_k$,
here $\delta_k$ may be a binning interval width,
the biased link entropy reads
\\[-2ex]
\hspace*{2em}
\begin{eqnarray}
H(\alpha)= - \displaystyle\sum_k 
\frac{k^\alpha p_k}{\delta_kN(\alpha)} \ln
\frac{k^\alpha p_k}{\delta_kN(\alpha)}.
\end{eqnarray}
\vspace*{-3ex}
\subsection{Node degree correlations: Entropy approach?
}
\vspace*{-.5ex}
The idea now is to use entropies instead of correlations or rank
correlations.
Naively one would use 
define an entropy
of all coefficients of the
node degree correlation matrix $p_{kl}$,
$H=-\sum_{kl} p_{kl} \ln p_{kl}$.
However,
then any invariances like
 $(k_1,k_2) \rightarrow (2k_1,2k_2)$ 
or  $(k_1,k_2) \rightarrow (k_0+k_1,k_0+k_2)$
are lost, 
but
such invariances would be desired for different
description levels of the systems.
Annother possible approach could be via the
Kullback-Leibler Distance
$D({p^{\rm A},p^{\rm B}}) = 
\sum_i p^{\rm A}_i \ln
({p^{\rm A}_i}/{p^{\rm B}_i}).
$
Here, one has to apply it to the node degree
$k_i^{\rm A}, k_i^{\rm B}$
for each link
$i$.
However, this is generically nonsymmetric
(for a symmetrized definition see \cite{johnson01}),
and again, there is 
 no invariance for
e.g.\
  $(k_1,k_2) \rightarrow (2k_1,2k_2)$.  
--- 
As a last approach, one could define a 
Biased Cross Link Entropy 
by replacing $k_i^{\rm B}$ by $k^\gamma \cdot k_i^{\rm B}$.
---
This discussion shows that simple definitions via link entropies bear difficulties.

\clearpage

\clearpage

\clearpage
\section{Definition of the Offdiagonal Complexity (OdC) \label{sec_ODCdef}}
\vspace*{-4ex}
Let $g_{ij}$ be the adjacency matrix of a graph 
with $N$ nodes, i.e., $g_{ij}=1$ if nodes $i$ and $j$ are
connected, else $g_{ij}=0$. 
   Then OdC is defined as follows \cite{claussenECMTB}.
\\
(i)
For each node $i$,
let $l(i)$ be the node degree, 
i.e.\ the number of edges (links),
\\[-2.5ex]
\begin{eqnarray}
l(i):= \sum_{j=0}^{N-1} g_{ij}
\end{eqnarray}
\mbox{}\\[-2.5ex]
(ii)
Let $c_{mn}$
be the number of edges between al pairs
of nodes $i$ and $j$,
with node degrees $m=l(i)$, $n=l(j)$
with $l(j)\geq l(i)$ (ordered pairs), i.e.,
\\[-2.5ex]
\begin{eqnarray}
c_{mn}:= 
\sum_{j=0}^{N-1}
\sum_{j=0}^{N-1} 
g_{ij}  \delta_{m,l(i)} \delta_{n,l(j)}
H(l(i)-l(j)).
\end{eqnarray}
\mbox{}\\[-2.5ex]
Here $\delta$ is the Kronecker symbol and 
$H(x)=1$ for $x\leq 0$ and $H(x)=0$ for $x<0$.
Due to the pair odering, the matrix $c_{mn}$ has entries only on
the main diagonal and above.
Thus, $c_{mn}$
is a (not normalized) node-node link correlation matrix.
 \\
\noindent 
(iii)
Summation over the 
minor diagonals, or offdiagonals,
i.\ e.\ 
all pairs with same 
$k_i-k_j$
up to $k_{\rm max}=\min_{i}\{l(i)\}$,
 and normalization,
\begin{eqnarray}
\tilde{a}_{k}= \sum_{i=0}^{k_{\rm max}-k} c_{i,k+i},
~~~~~~~~~~
A:=\sum_{k0}^{k_{\rm max}} \tilde{a}_{k},
~~~~~~~~~~
\forall_k a_k:= \tilde{a}_{k} / A.
\end{eqnarray}
 \\
(iv)
Then OdC is defined as an entropy measure on this 
normalized distributions (here it is understood that $0\ln(0)=0$),
\\[-1.5ex]
\begin{eqnarray}
{\rm OdC } =-\sum_{k=0}^{k_{\rm max}} a_k \ln a_k.
\end{eqnarray}

\noindent
OdC is an approximative complexity estimator
that takes as values
 zero for a regular 
lattice
(an orthogonal $n$-dimensional lattice with periodic boundaries 
consists of bulk nodes with
$2n$ neighbors. Thus $c_{2n,2n}=1$ is the only entry; 
for this regular structure OdC vanishes.
Also a 2-dimensional hexagonal lattice has only one entry),
 zero for a fully connected graph,
 low values for a random graph,
and
 higher values for `apparently complex' structures.
One main advantage is that it does not involve 
costly (high-order or NP-complete) computations.

\clearpage

\section{Application to the Helicobacter pylori
protein interaction graph and reshuffling to a
random graph\label{sec_helico}}
\begin{figure}[htbp]
\centerline{
 \epsfig{file=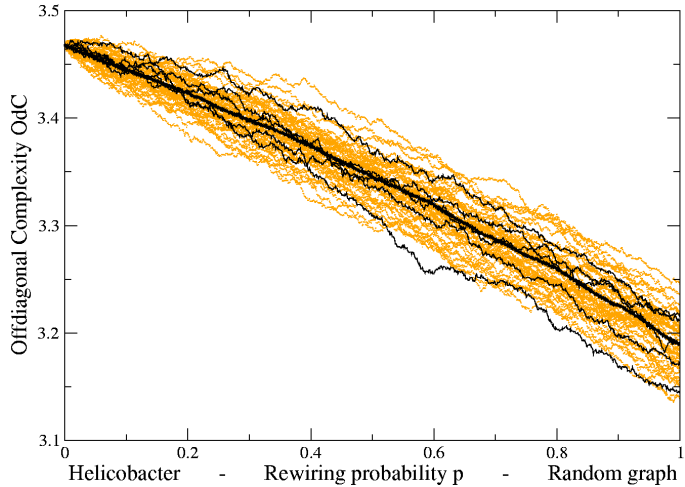,height=6.5cm}
}
\caption{OdC for random reshufflings of the {\sl Helicobacter pylori}
network (left, $p=0$) up to a rewiring probability of $p=1$ (right). 
The bold line shows the average, five OdC trajectories
along a rewiring path are shown for illustration (thin lines).
\label{helico}
}
\end{figure}

\vspace*{-2ex}
To demonstrate that OdC can distinguish between random
graphs and complex networks,
the 
Helicobacter pylori protein interaction graph 
\cite{helico_dat}
has been chosen.
For different rewiring probabilities $p$ and  $10^2$ realizations 
each, the links have been reshuffled, ending up with a random 
graph for $p=1$.
As can be seen in Fig.~\ref{helico},
rewiring in any case lowers the Offdiagonal Complexity.

\vspace*{-2ex}
\section{Conclusions and Outlook}
\vspace*{-2ex}
A new complexity measure for 
graphs and
networks has been proposed.
The motivation of its definition 
is twofold:
One observation is that the
binning of link distributions is
problematic for small networks.
Herefrom the second observation is that 
if one uses instead of the (plain) 
entropy of link distribution,
which is unsignificant for scale-free networks,
a ``biased link entropy'', it has an extremum where the
exponent of the power law is met.
\\
The central idea of OdC is to apply an entropy
measure to  
the
degree
correlation
matrix,
after summation over the offdiagonals.
This allows for a quantitative,
yet still approximative, measure 
of complexity.
OdC roughly is  `hierarchy sensitive'
and has the main advantage 
of being
computationally not costly.
\vspace*{-2ex}
\paragraph*{Acknowledgments.} 
\noindent
J.C.C.\ thanks  Christian Starzynski  for providing the  simulation code for Fig.~\ref{helico}, and an anonymous referee for  constructive remarks.

\end{document}